\documentclass[runningheads]{llncs}
\usepackage[T1]{fontenc}
\usepackage{graphicx}
\usepackage{color}
\usepackage{textcomp}

\usepackage{graphics}
\usepackage{graphicx}
\usepackage{booktabs}
\usepackage{tabularx}
\usepackage{float}

\usepackage{amssymb}
\usepackage{times}
\usepackage{epsfig}
\usepackage{comment}
\usepackage{amsmath}
\usepackage{amssymb}
 \usepackage{booktabs}
 \usepackage{multirow}
\usepackage{pifont}
\usepackage{caption}
\usepackage{subcaption}
\usepackage{url}
\usepackage{soul}
\usepackage{dirtytalk}
\usepackage[normalem]{ulem}
\usepackage[table,xcdraw]{xcolor}
\usepackage{hyperref}
\hypersetup{
    colorlinks=true,
    linkcolor=black,
    filecolor=black, 
    citecolor=black,
    urlcolor=blue,
}
\usepackage{xcolor}
\begin{document}
\title{FAU-Net: An Attention U-Net Extension with Feature Pyramid Attention for Prostate Cancer Segmentation}
 \author{Pablo Cesar Quihui-Rubio\inst{1} , Daniel Flores-Araiza \inst{1}, Miguel Gonzalez-Mendoza\inst{1}, and Christian Mata\inst{3}\inst{,4}, Gilberto Ochoa-Ruiz\inst{1} }
% %
 \authorrunning{P. Quihui-Rubio et al.}
% % (feature abused for this document to repeat the title also on left hand pages)

% % the affiliations are given next; don't give your e-mail address
% % unless you accept that it will be published
\institute{Tecnologico de Monterrey, School of Engineering and Sciences, Mexico. 
 \and Universitat Politècnica de Catalunya, 08019 Barcelona. Catalonia, Spain.
 \and Pediatric Computational Imaging Research Group, \\ Hospital Sant Joan de Déu, Esplugues de Llobregat, 08950, Catalonia, Spain \\}

\maketitle              % typeset the header of the contribution
\begin{abstract}
This contribution presents a deep learning method for the segmentation of prostate zones in MRI images based on U-Net using additive and feature pyramid attention modules, which can improve the workflow of prostate cancer detection and diagnosis. The proposed model is compared to seven different U-Net-based architectures. The automatic segmentation performance of each model of the central zone (CZ), peripheral zone (PZ), transition zone (TZ) and Tumor were evaluated using Dice Score (DSC), and the Intersection over Union (IoU) metrics. The proposed alternative achieved a mean DSC of 84.15\% and IoU of 76.9\% in the test set, outperforming most of the studied models in this work except from R2U-Net and attention R2U-Net architectures.
\end{abstract}
\section{Introduction}
% Definir Cancer de Prostata y su importancia
Prostate cancer (PCa) is the most common solid non-cutaneous cancer in men and is among the most common causes of cancer-related deaths in 13 regions of the world \cite{yongkai2018}.

When detected in early stages, the survival rate for regional PCa is almost 100\%. In contrast, the survival rate when the cancer is spread to other parts of the body is of only 30\% \cite{astrazeneca_2020}.
% Cómo se diagnostica?
Magnetic Resonance Imaging (MRI) is the most widely available non-invasive and sensitive tool for detection of PCa, due to its high resolution, excellent spontaneous contrast of soft tissues, and the possibility of multi-planar and multi-parametric scanning \cite{Chen_2008}.
Although MRI is used traditionally for staging PCa, it can be also be used for the PCa detection through the segmentation of Regions of Interest (ROI) from MR images.

%Relevancia de Segmentation de prostata
The use of image segmentation for PCa detection and characterization can help determine the localization and the volume of the cancerous tissue \cite{Haralick_1985}. This highlights the importance of an accurate and consistent segmentation when detecting PCa.

However, the most common and preferred method for identifying and delimiting prostate gland and prostate regions of interest
%(central zone, peripheral zone, and transition zone)%
is by performing a manual inspection by radiologists \cite{aldoj2020}.
% Desventajas de la segmentacion manual
This manual process is time-consuming, and is sensitive to specialists' experience, resulting in a significant intra- and inter-specialist variability. \cite{Rasch_2011}. 
%Proposal
Automating this process for the segmentation of prostate gland and regions of interest, in addition to saving time for radiologists, can be used as a learning tool for others and have consistency in contouring \cite{Mahapatra_2014}. 

%Relevance of deep learning
Deep Learning (DL) base methods have recently been developed to perform automatic prostate segmentation \cite{ELGUINDI_2019}. One of the most popular methods is U-Net \cite{Unet_ronneberger_2015}, which has been the inspiration behind many recent works in literature.

In this work, we propose an automatic prostate zone segmentation method that is based on an extension of Attention U-Net that combines two types of attention, pyramidal and additive. We also include the pixel-wise estimation of the uncertainty.

The zones evaluated in this work are the central zone (CZ), the peripheral zone (PZ), transition zone (TZ), and, in the case of a disease, the tumor (TUM), different from other works, which only evaluate CZ and PZ \cite{yonkai2020uncertainty}.

% Secciones de paper
the rest of this paper is organized as follows: Section \ref{sec:State_of_the_art} describes previous works dealing with the prostate segmentation. Section \ref{sec:methods} describes the dataset used in this work, the proposed architecture, as well as the experimental setup to evaluate it. In section \ref{sec:results} the results of the experiments are presented and discussed and Section \ref{sec:conclusion} concludes the article.

\section{State-of-the-Art}
\label{sec:State_of_the_art}
In medical imaging, one of the best known DL models in the literature for segmentation is U-Net,  which consists of two sub-networks: an encoder with a series of four convolutions and max-pooling operations to reduce the dimension of the input image and to capture its semantic information at different levels. The second sub-network is a decoder that consists of four convolution and up-sampling operations to recover the spatial information of the image \cite{Unet_ronneberger_2015}. The work from Zhu et \textit{al.} \cite{Zhu_deeplycnn_2017} proposes a U-Net based network to segment the whole prostate gland, obtaining encouraging results. Moreover, this architecture has served as the inspiration for some variants that enhance the performance of the original model. One example is the work from Oktay et \textit{al.} \cite{Att-unet-2018}, which proposes the addition of attention gates inside the original U-Net model with the intention of making the model focus on the specific target structures. In this architecture, the attention layers highlight the features from the skip connections between the encoder and the decoder. Many others extension architectures have been proposed since U-Net was released, some of them include Dense blocks \cite{dense_unet_wu_2021}, residual and recurrent blocks \cite{r2unet}, even novel architectures implemented transformers blocks named Swin blocks in order to obtain Swin U-Net \cite{swinunet}. 

All the mentioned models had demonstrated great results in many biomedical image datasets. However, in this work we focused on PCa segmentation, in particular, the main zones of the prostate, which has not been deeply investigated by some of these models. 

\section{Materials and Methods}
\label{sec:methods}
\subsection{Dataset}
\label{sec:dataset}

%This study was carried out in compliance with the \textit{Universidad Politécnica de Cataluña} (UPC) in Barcelona, and Centre Hospitalaire de Dijon in France.

This study was carried out in compliance with the Centre Hospitalier de Dijon. The dataset provided by these institutions consists of three-dimensional T2-weighted fast spin-echo (TR/TE/ETL: 3600 ms/ 143 ms/109, slice thickness:$1.25$ mm) images acquired with sub-millimetric pixel resolution in an oblique axial plane. The total number of patients from the dataset are 19, with a total of 205 images with their corresponding masks used as a ground truth.
The manual segmentation of each with four regions of interest (CZ, PZ, TZ, and TUMOR) was also provided, this process was cautiously validated by multiple professional radiologists and experts using a dedicated software tool \cite{Mata_2022,Mata_2020}. 

The entire dataset contains four different combination of zones, being: (CZ+PZ), (CZ+PZ+TZ), (CZ+PZ+Tumor), and (CZ+PZ+TZ+Tumor) with 73, 68, 23, and 41 images respectively. For the purpose of this work, the dataset was divided in 85\% for training and 15\% for testing, keeping a similar distribution in both sets of data, having a total of 174 images for training, and 31 for testing. 

\begin{figure}[htp]
    \centering
    \begin{subfigure}[t]{0.21\textwidth}
        \centering
        \includegraphics[width=1\textwidth]{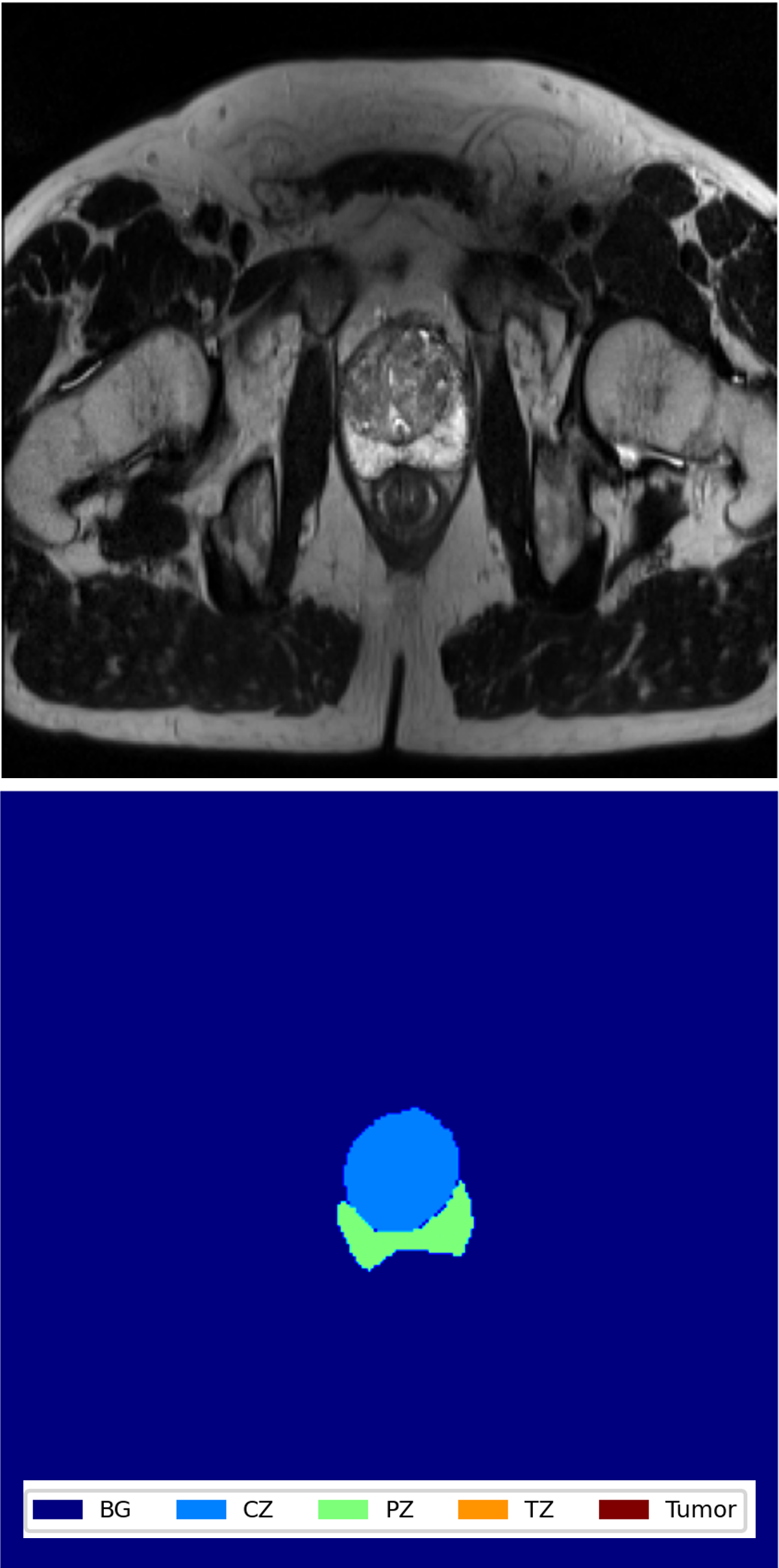}
        \caption{CZ and PZ}
        \label{fig:czpz}
    \end{subfigure}
    \hfill
    \begin{subfigure}[t]{0.21\textwidth}
        \centering
        \includegraphics[width=1\textwidth]{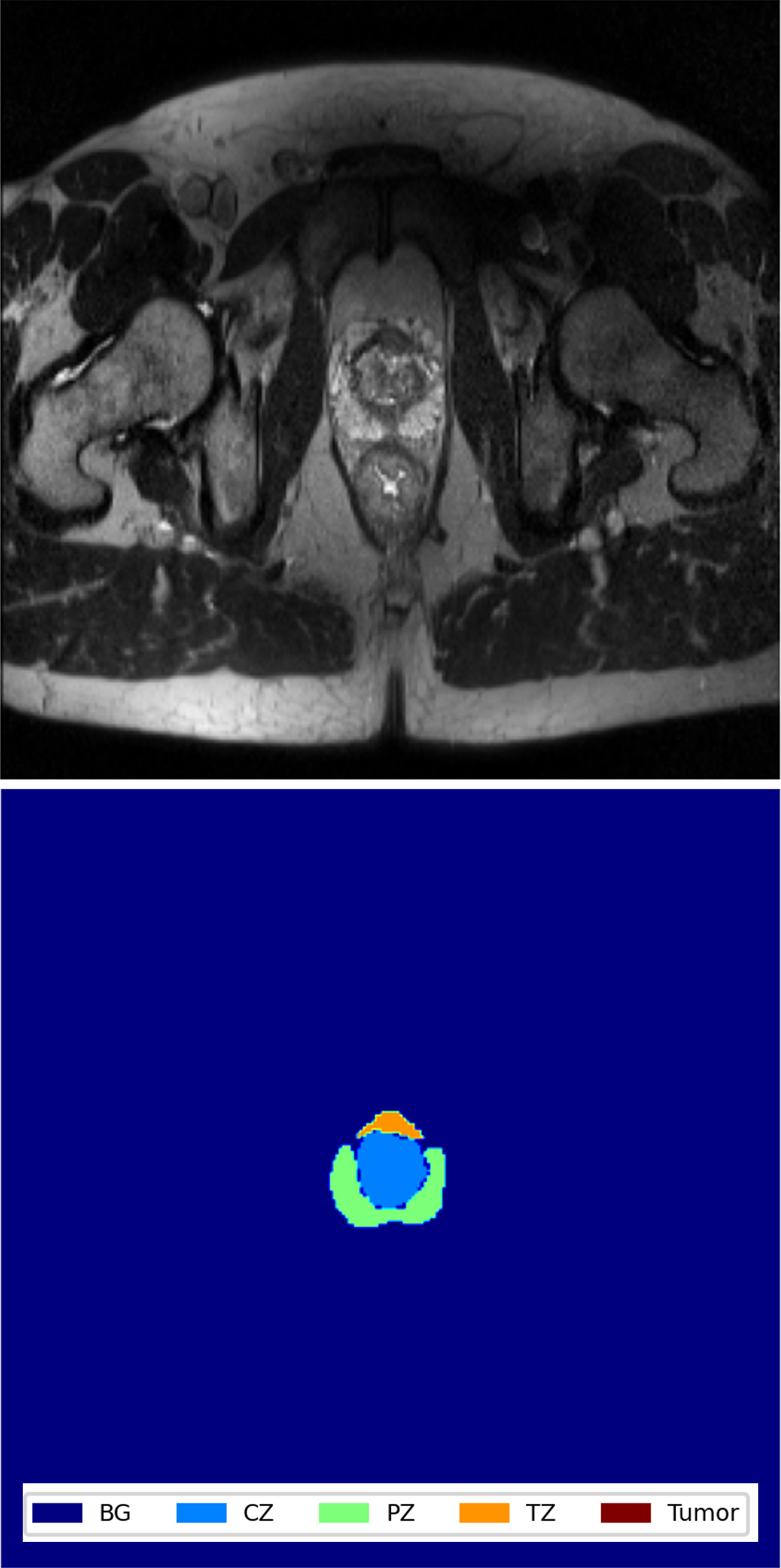}
        \caption{CZ, PZ, and TZ}
        \label{fig:czpztz}
    \end{subfigure}
    \hfill
    \begin{subfigure}[t]{0.21\textwidth}
        \centering
        \includegraphics[width=1\textwidth]{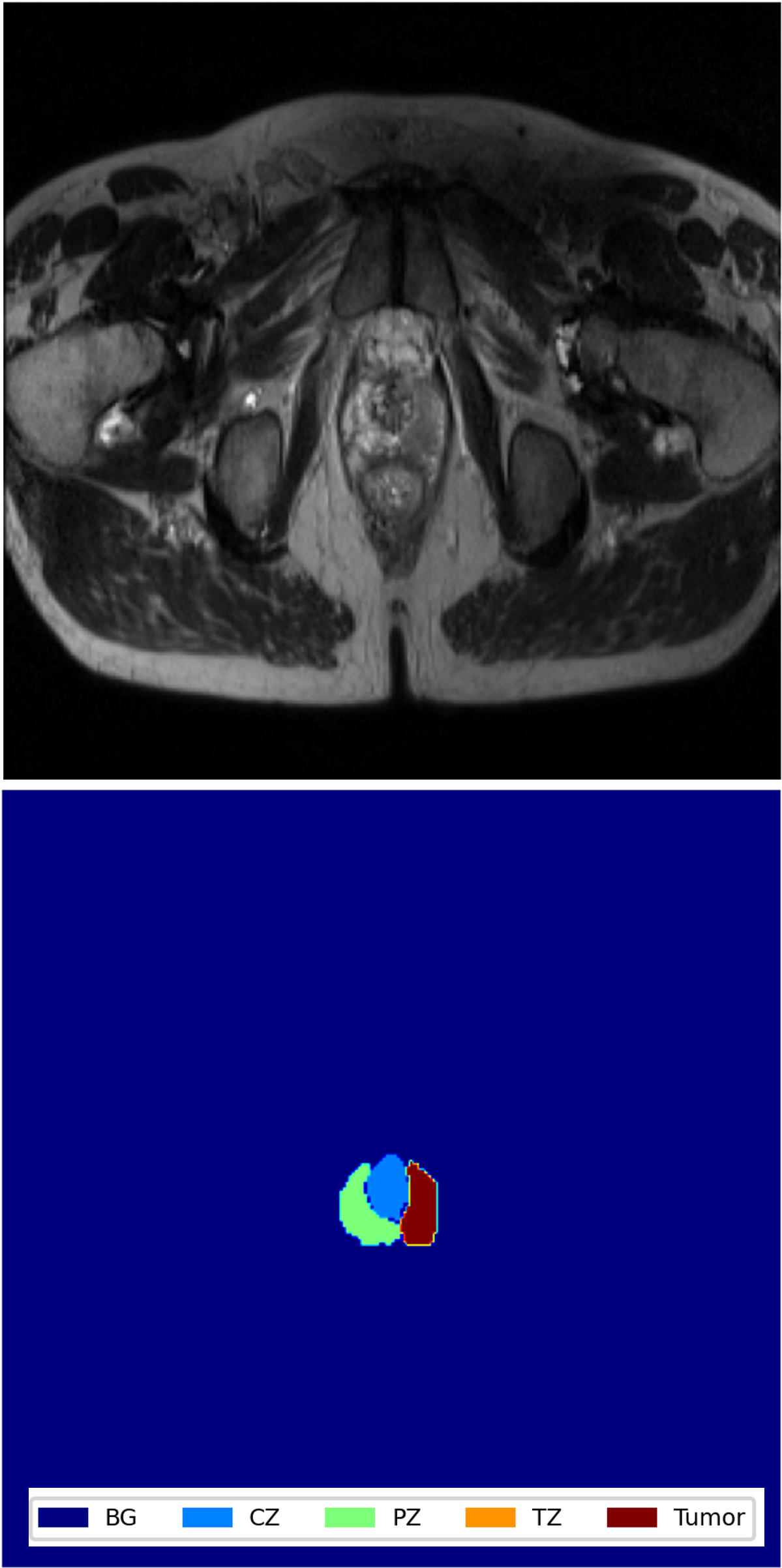}
        \caption{CZ, PZ, and Tumor}
        \label{fig:czpztum}
    \end{subfigure}
    \hfill
    \begin{subfigure}[t]{0.21\textwidth}
        \centering
        \includegraphics[width=1\textwidth]{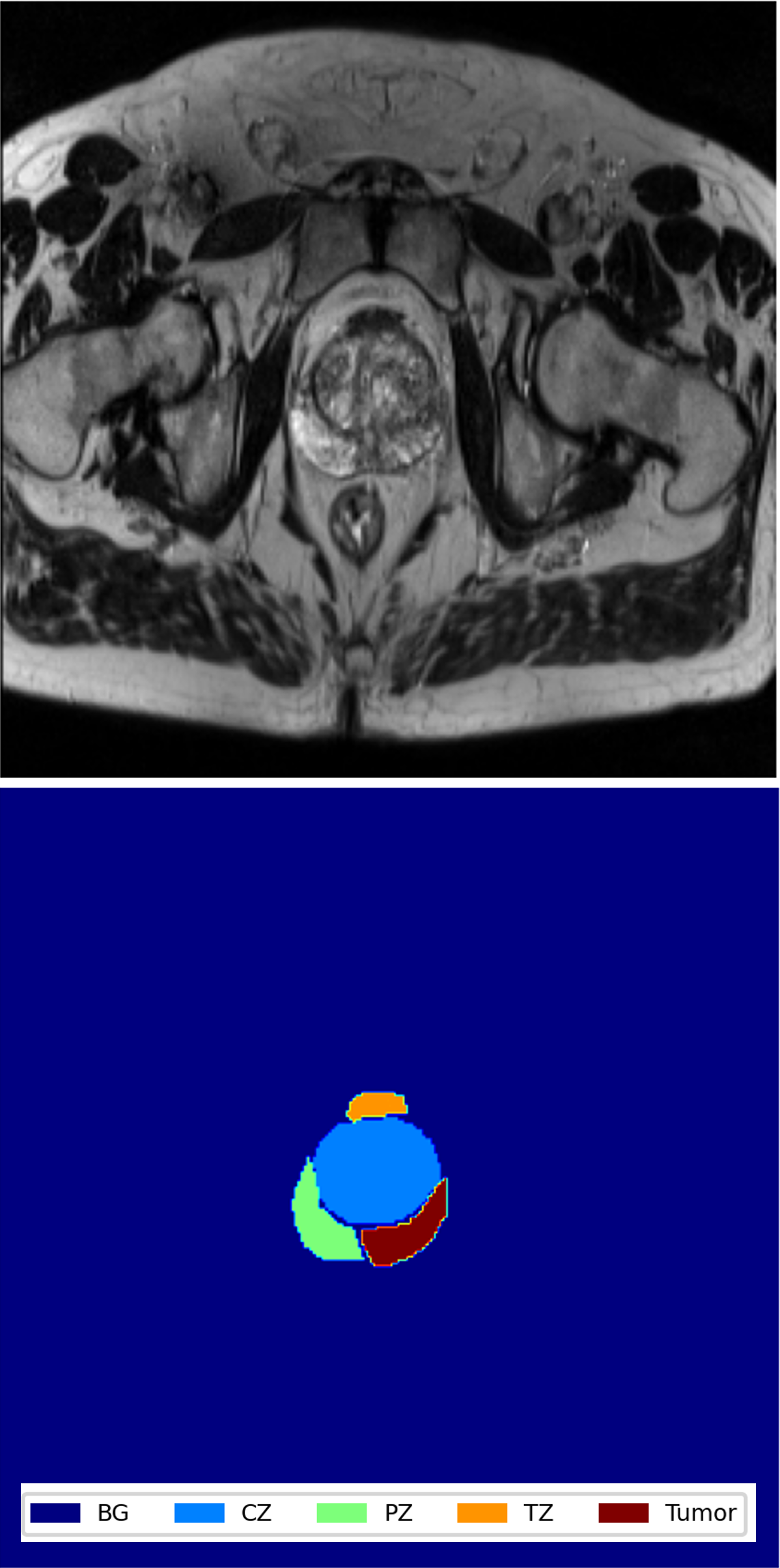}
        \caption{CZ, PZ, TZ, and Tumor}
        \label{fig:allzones}
    \end{subfigure}
    \hfill
   \caption{Sample images from every possible combination of zones in the dataset are presented in the upper row. Their respective ground truth masks are shown in the lower row.}
    \label{fig:dataset_examples} 
\end{figure}

In figure \ref{fig:dataset_examples} examples of images from every possible combination of zones in the dataset are presented.

\subsection{Feature Pyramid Attention}
The work of Yonkai et \textit{al.} \cite{yongkai2018} introduces the feature pyramid attention (FPA) network to capture information at multiple scales. It contains three convolutional blocks of different sizes ($3x3$, $5x5$ and $7x7$) to extract the features from different scales. These are then integrated from smaller to bigger, to incorporate the different scales. In our work, the attention map is multiplied by the features from the skip connection after a $1x1$ convolution. A visual representation of this attention block is presented in Figure \ref{fig:pyramid_attn}. 

\subsection{Proposed Work}
This contribution proposes the Fusion Attention U-Net (FAU-Net), an Attention-U-Net-based extension with pyramidal and additive attention. The proposed model is used to perform the segmentation of five different regions from the PCa dataset described in \ref{sec:dataset}.

\begin{figure}[htp]
\centering
\includegraphics[width=0.95\linewidth]{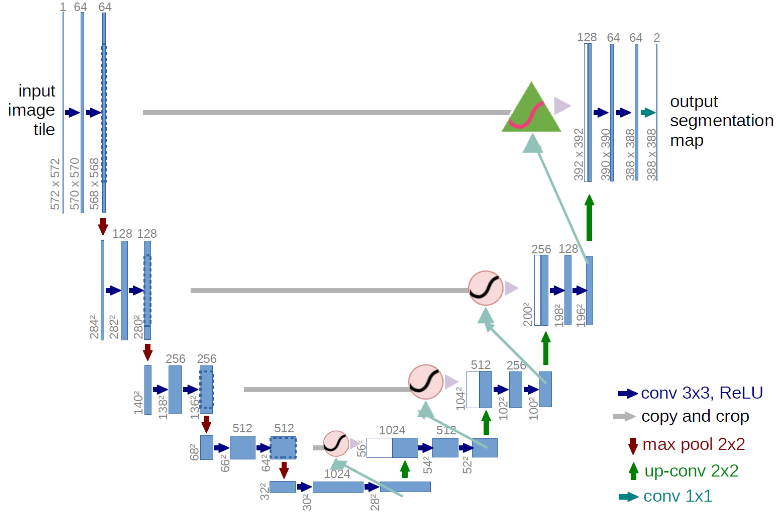}
\vspace{-0.15cm}
\caption{Proposed Fusion Attention U-Net model. The input image first goes through the contracting path. The boxes represent the feature map at each layer, and the blue boxes represent the cropped feature maps from the contracting path.}
\label{fig:unet}
\end{figure}

\begin{figure}[htp]
\centering
\includegraphics[width=0.85\linewidth]{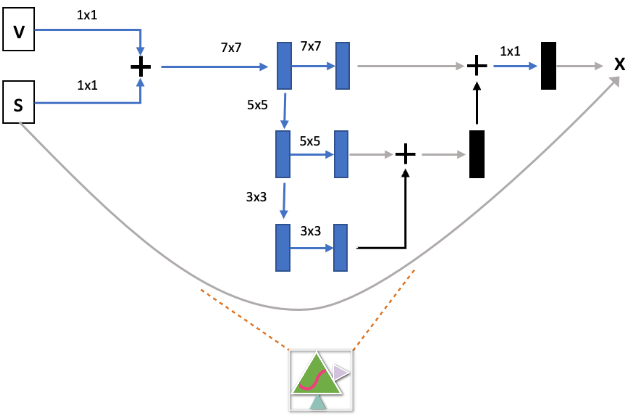}
\vspace{-0.15cm}
\caption{The feature pyramid attention block. It consists of three convolutional blocks of $3x3$, $5x5$, and $7x7$ which responses are integrated to capture the context of each level.}
\label{fig:pyramid_attn}
\end{figure}

Attention U-Net implements attention gates (AG) into the U-Net architecture to highlight salient features that are passed through the skip connections, these gates allow the network to disambiguate irrelevant and noisy responses in skip connections, leaving only the relevant activations to merge \cite{Att-unet-2018}. In the architecture proposed, we used AGs in the last three levels of the architecture. Meanwhile, in the first level, the implementation of a FPA was carried out to give further attention in those layers, were more data could be leaked as shown in Figure \ref{fig:pyramid_attn}.

A comparison between U-Net \cite{Unet_ronneberger_2015}, Attention U-Net \cite{Att-unet-2018}, Dense U-Net \cite{dense_unet_wu_2021}, Attention Dense U-Net \cite{Att-denseunet-2019}, R2U-Net \cite{r2unet}, Attention R2U-Net, Swin U-Net \cite{swinunet} and the proposed FAU-Net was done to validate the results obtained.

Most of the works in the literature perform the segmentation task of only two zones, and the number of works that consider a third zone (TZ) is limited, mainly because the boundaries are more delimited than zones such as TZ or Tumor. In this work we used a private dataset which incorporates the TZ and, in some cases, a tumor. This zone is important because it could lead to a proper diagnosis or treatment if a tumor is present.  

Therefore, we proposed an attention-based model to perform segmentation with a dataset of only T2-weighted images with 4 prostate zones, and compare the results against other models proposed in the literature. We analyzed the segmentation of the prostate zones using different metrics to choose the best DL architecture. Finally, we did a qualitative analysis of the predictions of each model. 

\begin{table}[htp]
\centering
\caption{Count of trainable parameters for each model analyzed during this work.}
\begin{tabular}{l|c}
\toprule
 Model & Number of parameters \\
\midrule
U-Net & 1,940,885 \\
Attention U-Net & 1,995,409 \\
\textbf{FAU-Net} & \textbf{2,158,505} \\
Dense U-Net & 4,238,389 \\
Attention Dense U-Net & 4,271,521 \\
R2U-Net & 6,003,077 \\
Attention R2U-Net & 6,036,081 \\
Swin U-Net & 26,598,344 \\
\bottomrule
\end{tabular}
\label{tab:parameters_seg}
\end{table}

In Table \ref{tab:parameters_seg} is shown the number of parameters, which are different for each model, being the one with the lowest number the original U-Net, and the Swin U-Net with the highest number of parameters. FAU-Net has only around 160,000 more parameters than U-Net and Attention U-Net, being the third model with less parameters.

All the models were trained with the same dataset for 145 epochs, using Adam optimizer with a learning rate of $0.0001$, batch size of 6, and categorical cross-entropy as loss function. The performance was evaluated using F1-score and Intersection over Union (IoU) as the main metrics. All the training was done using a NVIDIA DGX workstation, using a V100 GPU.

\section{Results and Discussion}
\label{sec:results}
The results of this work are divided in two subsections for further analysis and comparison between the models: quantitative and qualitative.

\subsection{Quantitative Results}
Table \ref{tab:results_seg} shows a summary of results for the evaluation of the eight studied architectures in two metrics (DSC and IoU) and loss value. Each evaluation corresponds to the mean value of the metrics for all the prostate zones and images in the test set. The bold values represent the model that achieved the best metric score within all of them. 

\begin{table}[htp]
\centering
\caption{The model performance evaluation was conducted using the Categorical Cross-Entropy (CCE) as the loss function. The metrics were designated with either an upward ($\uparrow$) or downward ($\downarrow$) arrow to indicate whether higher or lower values were desirable. Bold values and green highlights denote the best metric score achieved among all models.}
\label{tab:results_seg}
\begin{tabular}{@{}l|c|c|c@{}}
\toprule
\multicolumn{1}{c|}{Model} & IoU $\uparrow$    & DSC $\uparrow$    & Loss $\downarrow$\\ \midrule
U-Net                      & 70.76 & 80.00 & 0.0138 \\
Dense U-Net                & 74.53 & 83.65 & 0.0225 \\
Swin U-Net                 & 75.24 & 83.91 & 0.0124 \\
Attention U-Net            & 74.92 & 84.01 & 0.0114 \\
Attention Dense U-Net      & 75.12 & 84.01 & 0.0211 \\
FAU-Net                    & 75.49 & 84.15 & \cellcolor[HTML]{D8FFD8}\textbf{0.0107} \\ 
R2U-Net                    & 76.60 & 85.30 & 0.0131 \\
Attention R2U-Net          & \cellcolor[HTML]{D8FFD8}\textbf{76.89} & \cellcolor[HTML]{D8FFD8}\textbf{85.42} & 0.0120 \\
\bottomrule
\end{tabular}
\end{table}

As expected, the extended U-Net architectures performed better than the original U-Net architecture. For instance, the Dense U-Net model showed an improvement of approximately 5\% in both metrics. However, the Swin U-Net model, based on Swin Transformers and considered one of the best architectures available, did not perform as well on the dataset used in this study. It outperformed U-Net and Dense U-Net models in both metrics by 6\%, and Attention U-Net and Attention Dense U-Net in the IoU metric by only 0.4\% and 0.1\%, respectively. The subpar performance of this model could be attributed to various factors, but the most likely explanation is the small size of the dataset and the high number of training parameters, which may have led to overfitting.

Incorporating attention modules into U-Net and Dense U-Net models resulted in significant improvements compared to models without them. Attention U-Net outperformed U-Net by more than 5\% in both metrics. Meanwhile, Attention Dense U-Net achieved the same DSC score as Attention U-Net and a higher IoU score by approximately 1\%. These results indicate that attention modules are beneficial for obtaining better prostate segmentation, even with a relatively small dataset.

The proposed FAU-Net architecture in this study incorporated two types of attention: additive attention, as used in previous models, and pyramidal attention, consisting of attention modules in a cascade fashion. The objective of this model was to focus on the most complex features of each prostate image and obtain better information, and the results support this hypothesis. FAU-Net achieved IoU and DSC values of 75.49\% and 84.15\%, respectively, improving U-Net results by more than 6\%. However, this architecture was surpassed by R2U-Net and Attention R2U-Net.

R2U-Net and Attention R2U-Net are architectures that rely on recurrent residual blocks, which aid in extracting more information from deeper image features. In this study, Attention R2U-Net was the top-performing model overall, achieving metric scores greater than 76\% for IoU and 85\% for DSC, with a loss value of 0.0120.

\begin{figure}[htp]
\centering
\includegraphics[width=1\linewidth]{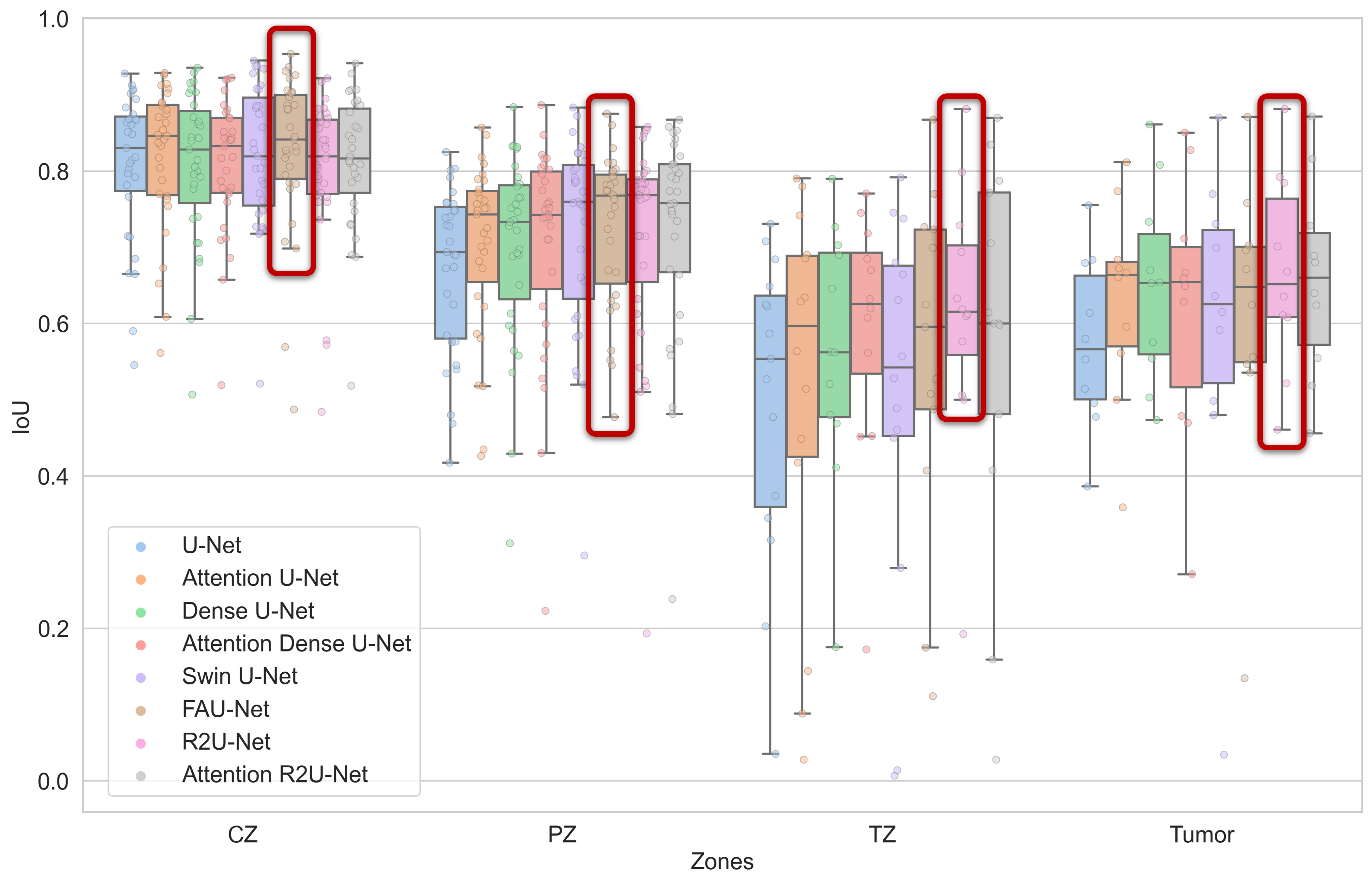}
\caption{The IoU scores obtained for each prostate zone from all images in the test set were compared between models. A line represents the median uncertainty value obtained, dots represent the particular score for each image, and the best model for each zone is indicated with a red box.} 
\label{fig:iou_boxplot}
\end{figure}

To gain a comprehensive understanding of the segmentation metrics in biomedical images, particularly related to the prostate, it is important to examine specific tissue zones. After analyzing the segmentation metrics through the full test set from the dataset, Figure \ref{fig:iou_boxplot} shows the IoU scores obtained from each image in each prostate zone. Each model is represented by a different color, and each test image is represented by a colored dot with the corresponding value. However, it's essential to note that not all images in the set had the same distribution, resulting in fewer dots in the boxplot for prostate zones such as TZ and Tumor. Nonetheless, the performance trends of the models in each particular zone can be analyzed.

Undoubtedly, the central and peripheral zones are the easiest for all models to segment, with only a few images having low IoU values. However, segmenting the peripheral zone appears slightly more challenging, likely due to its smaller size. The proposed FAU-Net was the best model overall for these two zones, with a mean IoU score of 82.63\% and 72.55\% for CZ and PZ, respectively. In contrast, the worst model was U-Net, with values below 80\% for CZ and 67\% for PZ.

As for the transition zone and tumors, the variation between the models is more noticeable in Figure \ref{fig:iou_boxplot}. Most models had lower values for outliers in the transition zone, achieving mean IoU scores lower than 60\% in all of them except R2U-Net, which managed to reach a mean score of 61\% in TZ.

Prostate tumors are a challenging task for segmentation due to the different types of geometry and boundaries between other tissues or zones. However, unlike TZ, most of the models managed not to have many outliers when segmenting the tumor, and most reached values higher than 60\%. The worst model for segmenting the tumor was U-Net, with a mean IoU score of only 57\%. On the other hand, the best model, R2U-Net, surpassed this model by 10\%, obtaining a mean IoU score of 67\%.

\subsection{Qualitative Results}

A visual inspection was carried out of the segmentation results of the eight models discussed in this study. This analysis of results complements the previous quantitative analysis based on the metrics. In this inspection, the images from the test set were visually compared to their corresponding ground truth, and conclusions were stated.

\begin{figure}[htp]%[htp]
    \centering
    \includegraphics[height=0.9\textheight]{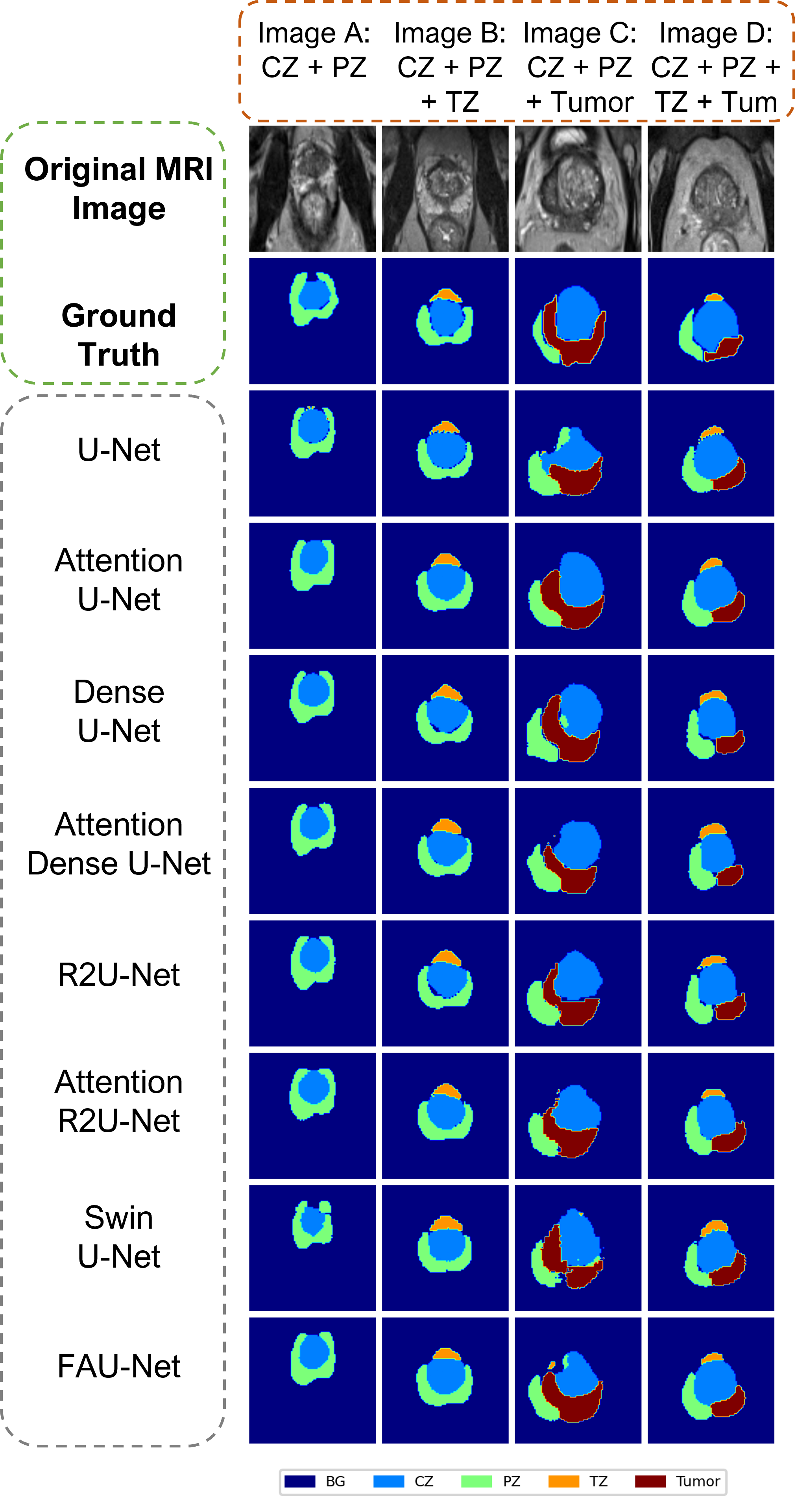}
    \caption{Image comparison of segmentation results using U-Net-like architectures. All possible combinations of zones available in the dataset are used as examples for conducting predictions on MRI images.}
    \label{fig:preds_imgs}
\end{figure}

Figure \ref{fig:preds_imgs} presents a qualitative comparison between each model's prediction in four different example images from the dataset, with all the possible combinations of zones. The first two rows show the original T2-MRI image of the prostate and below its corresponding ground truth. Then, each row represents prediction of the different models.

Starting from the base model U-Net, it is clear that U-Net had difficulty correctly segmenting all pixels, especially in images with tumors, for example in image C, this model missed many pixels that corresponded to the Tumor; this could be a wrong lead for a radiologist who is relying upon this model. Nevertheless, even though a Tumor is present in example D, U-Net segments most of the pixels better than in the previous example, at least from a visual perspective.

Based on qualitative analysis, some models, such as Attention U-Net, R2U-Net, and FAU-Net, performed better in segmenting all prostate zones, including the Tumor. Compared to the other models, these models produced smoother and more complete segmentation in images with three or more zones. However, it should be noted that FAU-Net misclassified some pixels as TZ in example C, which does not include TZ.

It is clear that images with only two zones (CZ and PZ) are easier to segment for all the models, which are the bigger and more present ones in the dataset. Some models in examples C and D include more pixels in the smaller zones, resulting in a smoother segmentation; although this looks great from visual analysis, compared to the ground truth, that prediction is incorrect; thus, relying solely on visual analysis is not advisable.

As a qualitative conclusion of the predictions based on the examples from Figure \ref{fig:preds_imgs}, it can be demonstrated that Attention U-Net and R2U-Net are the models with the best segmentation performance overall. However, based on the metrics and a visual analysis from the entire test set, in general the best performance was obtained by FAU-Net, R2U-Net, and Attention R2U-Net.

% For the segmentation task, not only the prediction results are important, but also being able to identify the uncertainty values since it could help determine whether or not a prostate was segmented correctly. In Figure \ref{fig:uq_maps} there are the same four example images and masks from the previous figure, and below are the uncertainty maps for each model after 50 predictions with the corresponding color bar for uncertainty value. The higher the value (red) more uncertainty in those pixels, where the smaller (blue) the model is more certain of that pixel prediction. 

% It is easy to identify that the model with higher uncertainty is U-Net in this examples, having more pixels closer to red than the other methods, which this higher values happen more often between the boundaries of prostate zones, mainly between the zones: CZ with TZ, and CZ with Tumor. However, Attention Dense U-Net also shows high values between the boundaries of the zones in the last example with the four zones. One more time, it can be demonstrated that all the models performed very well and with low uncertainty in images with only CZ and PZ.

% \begin{figure}[]
% \centering
% \includegraphics[width=0.8\columnwidth]{figures/faunet_uq.png}
% \caption{Comparison of uncertainty maps after 50 predictions for each model in all the combinations of zones in the dataset.} \label{fig:uq_maps}
% \end{figure}

\section{Conclusion}
\label{sec:conclusion}
In this work, we proposed a U-Net extension using two attention blocks: additive and pyramidal. From the results shown in Section \ref{sec:results}, we can conclude that the proposed architecture, FAU-Net, outperforms most of the studied architectures in this work. Moreover, other alternatives like R2U-Net and Attention R2U-Net, are still better suited to perform over this particular dataset than the proposed architecture. Furthermore, FAU-Net presents great metrics score and although it struggles in particular zones like TZ and Tumor, it is the best model to segment the CZ and PZ regarding the segmentation metrics in our dataset. 

Considering that the results obtained are promising, further investigation can be done by improving the FAU-Net architecture to achieve even better results. For instance, a future implementation of feature pyramid attention module in the R2U-Net architecture can lead to promising results using the dataset studied in this work for prostate segmentation. Also, trying more combinations of the attention modules and/or adding more levels to the architecture can produce interesting results.

\section{Acknowledgments}

The authors wish to acknowledge the Mexican Council for Science and Technology (CONACYT) for the support in terms of postgraduate scholarships in this project, and the Data Science Hub at Tecnologico de Monterrey for their support on this project. 

\bibliographystyle{splncs04}
\bibliography{references, references2}
\end{document}